\documentclass[aps,prl,amsmath,amssymb,preprint,superscriptaddress]{revtex4-1}%

\usepackage{graphicx}
\usepackage{dcolumn}
\usepackage{bm}

\begin{document}

\preprint{AIP/123-QED}

\title{Random lasing in structures with multi-scale transport properties}

\author{Marco Leonetti}
\affiliation{ISC-CNR, UOS Sapienza, P. A. Moro 2, 00185 - Roma, Italy}
\affiliation{Instituto de Ciencia de Materiales de Madrid (CSIC), Calle Sor Juana Ines de la Cruz 3, 28049 Madrid Espa\~{n}a.}

\homepage[]{www.luxrerum.org}\email[]{marco.leonetti@roma1.infn.it}

\author{Cefe Lopez}%
\affiliation{Instituto de Ciencia de Materiales de Madrid (CSIC), Calle Sor Juana Ines de la Cruz 3, 28049 Madrid Espa\~{n}a.}


\date{\today}

\begin{abstract}
In a random laser (RL), a system possessing in itself both resonator and amplifying medium while lacking of a macroscopic cavity, the feedback is provided by the scattering, which forces light to travel across very long random paths. Here we demonstrate that RL properties may be tuned by the topology of the scattering system retaining unchanged scattering strength and gain efficiency. This is possible in a system based on sparse clusters, possessing two relevant structural lengths: the macroscopic inter cluster separation and the mesoscopic intra-cluster mean free path.
\end{abstract}

\maketitle

The understanding of the interaction of light with complex systems is of paramount importance for the possibility to tailor optical properties and to realize the future light-driven photonic chip. There are many techniques that allow building nano-sized photonic structures, like direct laser writing \cite{Fischer:11}, ion beam milling \cite{0960-1317-11-4-301}, or electron beam lithography \cite{E_beam_lit} but in the last decades the self assembly of micron sized building blocks demonstrated to be between the most cost effective strategies for light science\cite{ADMA:ADMA201000356}. This approach allows the fabrication of large scale and cheap ordered and disordered photonic structures like photonic crystals \cite{1464-4258-8-5-R01, PhysRevA.83.023801}, and photonic glasses \cite{ADMA:ADMA200900827}.

The expertise in the fabrication of such structures allows now not only to tailor light propagation trough Bloch modes in an ordered structure but also to engineer the regime of light diffusion \cite{ADFM:ADFM200902008,Barthelemy2008} when disorder is prominent because the material's properties define the propagation characteristics. The overall scattering strength that is measured by the transport mean free path $\ell$, depends both on the size and on the spatial distribution of the scatterers which may vary by many orders of magnitude, like from colloidal structures to intergalactic dust\cite{1074123}.

If optical amplification is introduced in a disordered structure the balance between gain and losses brings about a lasing threshold that when surpassed, switches on a RL \cite{Wiersma_Rew}. In RL, that is the first lasing material possessing in itself both resonator and amplifying medium while lacking a macroscopic cavity, the feedback is provided by the scattering that confines the light in the amplifying volume for very long times forcing it to very long paths. Optimization of such RLs has been focused, up to now, on various aspects like gain efficiency \cite{Leonetti:09}, the scattering strength\cite{PhysRevA.74.053812}, and also on the nature of the building blocks \cite{Gottardo2008}. On the other hand the RL is a very complex phenomenon because both local (resonances hosted by microcavities of the structures \cite{PhysRevLett.82.2278}) and extended (the overall scattering efficiency \cite{PhysRevA.80.013833, Fallert_coexistence_nature, PhysRevLett.98.143901}) properties of the system concur in the emission process, while the role of the interplay between local and extended properties has been investigated just recently.

\begin{figure}
\centering
\includegraphics*[width=82.5 mm]{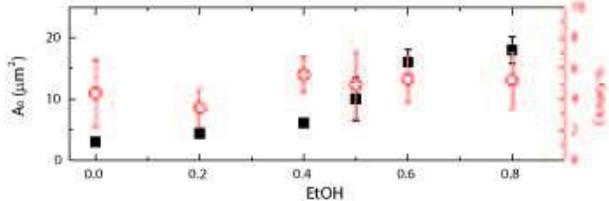}
\caption{The graph reports the A$_0$ parameter (squares, errors derived from the fit root mean squared error) and the average density (open circles) measured for samples with different ethanol concentration. Error bars for density has been obtained by statistical averages on 10 different portions measurements from different positions from five different samples with the same concentration. Absence of the bars indicate an error smaller than the marker size.}
\label{Fig1}
\end{figure}

Here we demonstrate that the improvement of the disordered lasing materials may be achieved on a system composed of clusters of titanium dioxide lying sparsely in a bath of dye-doped solution thus possessing two relevant lengths: the macroscopic inter cluster separation and the mesoscopic cluster size. It is possible to tune RL properties by changing the topology of the scattering system while retaining unchanged extensive properties: the scatterers density and the gain efficiency.

In practical terms the cluster size may be tuned starting from a titanium dioxide sol and by controlling the inter particle interaction by changing the ratio between polar (ethanol) and less polar (diethylene glycol, DEG) liquids composing the external phase\cite{Leonetti2011}.

\begin{figure}
\centering
\includegraphics*[width=82.5 mm]{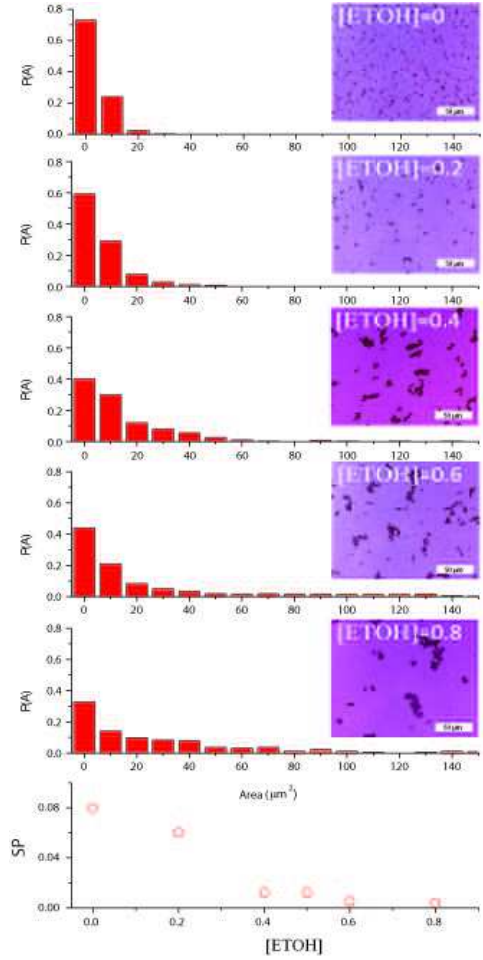}
\caption{The top five graphs report cluster size distributions for different values of [EtOH], ranging from 0 to 0.8. The last graph report the area fraction occupied by isolated particles as a function of the ethanol concentration [EtOH].}
\label{Fig2}
\end{figure}

The actual sample consists of an almost cylindrical volume of dye solution confined between two microscopy coverslips (separated by plastic spacers and laterally sealed by cyanoacrylic glue)  containing titania particles that self assemble on the bottom surface.  After one day the majority of the ethanol will be evaporated leaving clusters in a stable configuration embedded in a solution mainly composed of dye doped diethylene glycol. The size of clusters may be controlled by changing the ratio between ethanol and diethylene glycol in the starting solution without affecting the density (variations smaller than 2\%) and the amount of titanium dioxide present in the sample as shown in in Figure \ref{Fig1}. The area distribution of the clusters is non-gaussian as shown in figure \ref{Fig2}. The relative abundance exponentially decreases with size following the expression:

\begin{equation}
P(A)= k*exp(\frac{-A}{A_0}).
\end{equation}
We find that this is a good description of the size distribution since such a model fits the data and casts a size parameter A$_0$ with a small fitting error. In the graph of  \ref{Fig1} we report  A$_0$  (full squares scale on the left) as a function of the [EtOH] that has been probed in the range 0-0.8. With higher [EtOH] concentrations the samples has been found unstable (strong evaporation). Cluster area and cluster density has been evaluated by analyzing digitized microscope images (such as those  reported in the insets of figure \ref{Fig2}). Area has been obtained by counting pixels composing the clusters once the image is rendered in black and white (BW) by selecting an appropriate intensity threshold equal for all the measurements. Individual nanoparticles (structures with size much smaller than the average particle size) have not been taken into account in this analysis. Density has been estimated by calculating the ratio between the black (strongly scattering) and the white (transmitting) areas in the BW images. The clusters thickness (their vertical dimension) has been estimated by low depth of field optical microscope that furnished a value varying between some micrometers up to 25 $\mu$m.

Images of the samples with different ethanol concentration [EtOH] and obtained by a 20$\times$ optical microscope are reported in the insets of figure \ref{Fig2} together with graphs detailing the normalized cluster size distributions P(A) of the Area.  Agglomerates with size smaller than a micrometer are excluded from the statistics, because have high probability of being composed by a single particles (SP). Cluster size distributions are derived from the analysis of ten images from each sample. The last graph of the figure, reporting the ratio SP between the area occupied by single particles and the total area occupied by titanium dioxide as a function of [EtOH] confirms that the increasing presence of hydroxyl improves the clustering process.

The inter-particle interaction is critical in defining the stability of a colloid, \cite{ADMA:ADMA200900827,Book_1}. A key role is played by the particle surface charge that depends on the percentage and degree of polarity of liquids composing the solution. The surface of titania particles is capped with hydroxyl groups for which it is favorable to donate protons so that if placed in a medium unable to accept them the particles are unable to donate them and may even be forced to accept some, resulting in neutral or even positively charged particles. The addition of ethanol to the solution enables the particle to release protons and switches the surface charge to negative. The average effect of the presence of a polar liquid in the solution is to lower the zeta potential\cite{Chadwick2002229} thus favoring aggregation.

Having demonstrated that we can control the clusters' size we now turn to study the effects on the laser action that may be obtained upon external pumping. The experimental setup we employed is a system capable  to shine and generate population inversion in predefined, computer-designed area. This is achieved by exploiting a spatial light modulator (SLM) in amplitude modulation configuration: vertically polarized light from the pump laser is reflected by an SLM (model Holoeye LC-R 1080 1962x1200 pixels) in which each pixel changes by 90 $^\circ$ the polarization of the reflected light when set in the ON state. An image of the SLM is produced on the sample by using two lenses as described in reference\cite{Leonetti2011}. After the SLM the vertical polarization is selected by a reflective polarizer so that the activated pixels are not transmitted on the sample. With this approach which allows to achieve diffraction limited resolution, we will measure the effects of the anomalous diffusion on the lasing efficiency.

\begin{figure}
\centering
\includegraphics*[width=82.5 mm]{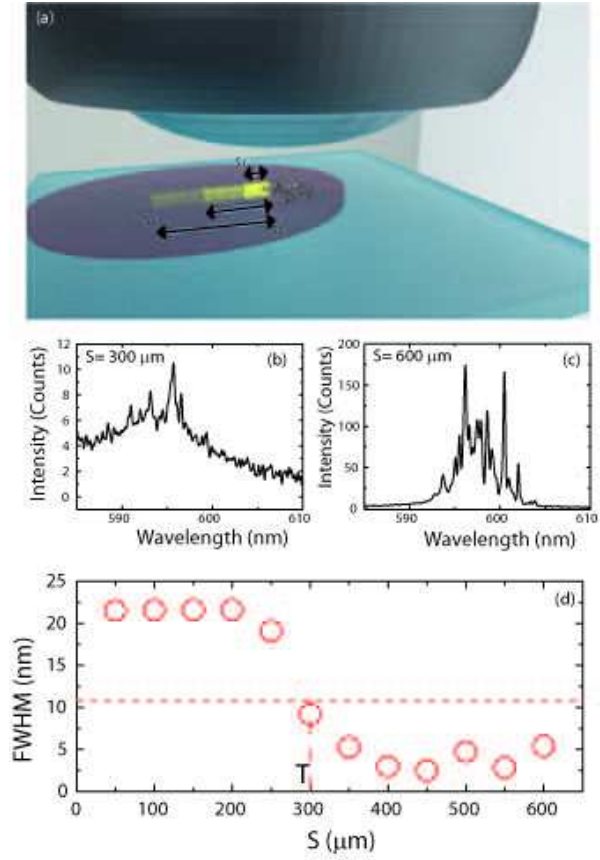}
\caption{(a) a sketch of the sample under pumping and of the collection objective (pumping optics are not shown as they lie below the sample). The stripe shaped inverted area may be of different sizes such as S1 S2 and S3 shown in the picture. In panel b) and c) the spectra (obtained with a 303 mm focal length spectrograph, 0.25 nm resolution) obtained from a single cluster placed at the edge of the stripe for small (b, S =300 $\mu m$) and large (c, S =600 $\mu m$) stripe length. In panel d) the full width at half maximum as a function of the stripe length S is reported. The horizontal dashed line indicates the threshold condition, while the vertical dashed line indicates the value of T.}
\label{setup}
\end{figure}

We will demonstrate in fact that the sizes distribution affects the overall efficiency of this lasing material. Indeed light traveling in the sample plane experiences a particular regime of propagation suffering a very strong scattering inside a cluster while traveling in straight line from one cluster to the next. Thus our system may be approximated as possessing two different relevant scattering mean free paths: an inter cluster mean free path $\ell_f$  and an intra-cluster mean free-path. An estimate of $\ell_f$  may be obtained directly from the microscope images. This is done  trough a computer routine that selects a single cluster and measures the distance from its first neighbors (counting the pixel separating the two clusters) in all directions. The operation is repeated for all the clusters in the image and the average of the set of distances obtained is $\ell_f$.

It has been shown recently that a particular distribution in the size of the particles composing a diffusive medium may give rise to diffusion regimes characterized by non Gaussian path length distributions. In particular in systems showing a power law distribution for mean free path, light propagation may be modeled in the framework of the Levy statistics \cite{ADFM:ADFM200902008, Barthelemy2008}. In our system instead the presence of an inter-particle space induces a twofold distribution of the mean free path, thus we will study the effects of $\ell_f$ on the lasing threshold which is one of the most important characteristics of a RL defining the efficiency of the balance between gain and losses.

To study effects of $\ell_f$ on lasing threshold we prepared a setup in which a single cluster is placed at one of the edges of a stripe shaped pump spot of thickness 10 $\mu$m and variable length S (see Fig. \ref{setup}). The stripe generates a flux of stimulated emission that is used as directional pump for the cluster \cite{Leonetti:11}. We make sure that the energy density over the whole stripe extension is kept constant at 0.5 $nJ/\mu m^2$. This is achieved by spatial filtering and magnification of the laser spot previous to the spatial modulation: in practice we exploit just the flattened top of the Gaussian profile in order to have a nearly homogeneous intensity on the reflective part of the SLM used for the spatial modulation.

Photons generated by stimulated emission in the intercluster space supply energy to the cluster. This is demonstrated in figure \ref{setup} panels b and c, lasing emission from a cluster  shined with S=300 $\mu m$ is less intense than the the emission from the same cluster but with a  pumping stipe of S=600 $\mu m$. We stress here that light is collected just from the cluster: no emission is retrieved when the optics is aligned to the inter-cluster space. However the path followed from light before of hitting the target cluster plays an important role: the chance that amplified stimulated emission has to reach the titanium dioxide structure is lowered by the scattering suffered along that path. Therefore in our system the inter cluster separation distribution affects directly the threshold. In the following we will demonstrate that the lasing efficiency directly depends on the $\ell_f$ parameter.

In standard experiments threshold T is estimated by measuring the narrowing of the full width at half maximum (FWHM) of emitted spectra as a function of the pump energy. A similar behavior for the FWHM is found if S is varied instead of energy, (see from figure \ref{setup} d). This happens because, as in variable stripe length experiments \cite{Leonetti:09, PhysRevA.83.023801}, the illuminated stripe act as an one dimensional amplifier at the edges of which amplified spontaneous emission (ASE) is channeled. Thus increasing the size S of the stripe results in an exponential increase of the ASE intensity pumping the cluster.

\begin{figure}
\centering
\includegraphics*[width=82.5 mm]{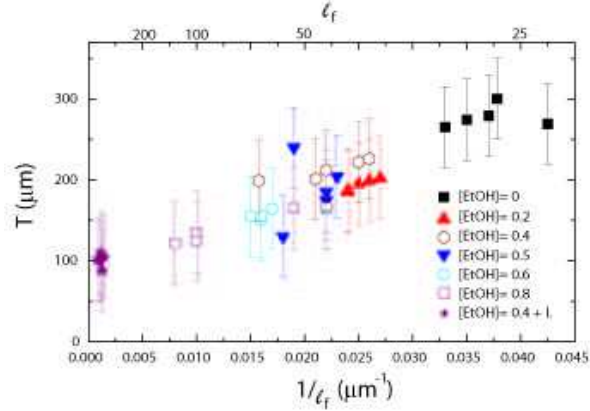}
\caption{Lasing threshold measured in samples with different $\ell_f$. Error bars result from the resolution of the threshold measurement. All measurements have been performed with an input fluence of  0.5 nJ/$\mu$m$^2$ per pulse. The samples indicated with "[EtOH]= 0.4 + I" ( full stars)  results from the isolation process.}
\label{Fig4}
\end{figure}

In our experimental protocol T is defined as the value of S, for which the full width is the average between its maximum and minimum values:
\begin{equation}
(Max(FWHM)-Min(FWHM))/2= FWHM(T)
\end{equation}
(where max and min represent respectively the maximum and the minimum values for all the FWHM obtained). We report T in figure \ref{Fig4} as a function of $\ell_f$. Five samples have been prepared for each [EtOH] value that, due to different disorder realization, show slightly different values of $\ell_f$. Each point results from the measurements of a randomly chosen cluster with a diameter between 10 and 15 $\mu$m. The figure shows a clear decrease of the lasing threshold as a function of the inter cluster transport mean free path. The last points of the graph (purple stars) has been obtained by performing the ``isolation procedure'' described in \cite{Leonetti_pra2011} that allows to isolate a single cluster by exploiting optically generated hydrodynamical fluxes. In practice the samples originally created with [EtOH]=0.4 possessing $\ell_f$ around 50$\mu m$ have been used to obtain nearly isolated clusters. In this case an estimate (lower limit) for the $\ell_f$ is given by the distance to the cluster to the one under examination that ranges between 800 and 1200 $\mu$m.

We also confirmed that the presence of a variable percentage of ethanol does not influence the efficiency of the gain molecules. We measured the gain length G by using the variable stripe length technique that allows to measure optical amplification in nanostructures \cite{PhysRevA.83.023801}. We retrieved G =100$\pm$4 cm$^{-1}$ for a solution containing only DEG and G=106$\pm$8 cm$^{-1}$ for a solution with DEG with [EtOH]=0.8.

To summarize we demonstrated that it is possible to control the topology of self assembling of titanium clusters by regulating the density of hydroxyl groups that sets the inter particle interaction. Samples in which clusters with different size distribution and embedded in a dye-doped solution have been fabricated. Then we proved that the inter-cluster mean free path affects the lasing efficiency. By acting on the fashion in which the building block of the disordered structure are arranged, the lasing threshold has been reduced and, even more striking, this has been done without changing the density of the scatterers or the gain of the active medium, thus effectively increasing the efficiency achievable from disordered lasing materials. This study is suitable for the fabrication of a cheap photonic material with increased emission capability and synthesizable at large scale. The future step for its improvement is the polymerization of the inter cluster space that will allow to fabricate solid and stable chip with controllable lasing efficiency.

\begin{acknowledgments}
We thank Marta Ibisate, for fruitful discussions. The work was supported by; EU FP7 NoE Nanophotonics4Enery Grant No 248855; the Spanish MICINN CSD2007-0046 (Nanolight.es); MAT2009-07841 (GLUSFA) and Comunidad de Madrid S2009/MAT-1756 (PHAMA).
\end{acknowledgments}

\bibliography{lpr-demo}{}

\begin{thebibliography}{23}%
\makeatletter
\providecommand \@ifxundefined [1]{%
 \@ifx{#1\undefined}
}%
\providecommand \@ifnum [1]{%
 \ifnum #1\expandafter \@firstoftwo
 \else \expandafter \@secondoftwo
 \fi
}%
\providecommand \@ifx [1]{%
 \ifx #1\expandafter \@firstoftwo
 \else \expandafter \@secondoftwo
 \fi
}%
\providecommand \natexlab [1]{#1}%
\providecommand \enquote  [1]{``#1''}%
\providecommand \bibnamefont  [1]{#1}%
\providecommand \bibfnamefont [1]{#1}%
\providecommand \citenamefont [1]{#1}%
\providecommand \href@noop [0]{\@secondoftwo}%
\providecommand \href [0]{\begingroup \@sanitize@url \@href}%
\providecommand \@href[1]{\@@startlink{#1}\@@href}%
\providecommand \@@href[1]{\endgroup#1\@@endlink}%
\providecommand \@sanitize@url [0]{\catcode `\\12\catcode `\$12\catcode
  `\&12\catcode `\#12\catcode `\^12\catcode `\_12\catcode `\%12\relax}%
\providecommand \@@startlink[1]{}%
\providecommand \@@endlink[0]{}%
\providecommand \url  [0]{\begingroup\@sanitize@url \@url }%
\providecommand \@url [1]{\endgroup\@href {#1}{\urlprefix }}%
\providecommand \urlprefix  [0]{URL }%
\providecommand \Eprint [0]{\href }%
\providecommand \doibase [0]{http://dx.doi.org/}%
\providecommand \selectlanguage [0]{\@gobble}%
\providecommand \bibinfo  [0]{\@secondoftwo}%
\providecommand \bibfield  [0]{\@secondoftwo}%
\providecommand \translation [1]{[#1]}%
\providecommand \BibitemOpen [0]{}%
\providecommand \bibitemStop [0]{}%
\providecommand \bibitemNoStop [0]{.\EOS\space}%
\providecommand \EOS [0]{\spacefactor3000\relax}%
\providecommand \BibitemShut  [1]{\csname bibitem#1\endcsname}%
\let\auto@bib@innerbib\@empty
\bibitem [{\citenamefont {Fischer}\ and\ \citenamefont
  {Wegener}(2011)}]{Fischer:11}%
  \BibitemOpen
  \bibfield  {author} {\bibinfo {author} {\bibfnamefont {J.}~\bibnamefont
  {Fischer}}\ and\ \bibinfo {author} {\bibfnamefont {M.}~\bibnamefont
  {Wegener}},\ }\href@noop {} {\bibfield  {journal} {\bibinfo  {journal} {Opt.
  Mater. Express}\ }\textbf {\bibinfo {volume} {1}},\ \bibinfo {pages} {614}
  (\bibinfo {year} {2011})}\BibitemShut {NoStop}%
\bibitem [{\citenamefont {Reyntjens}\ and\ \citenamefont
  {Puers}(2001)}]{0960-1317-11-4-301}%
  \BibitemOpen
  \bibfield  {author} {\bibinfo {author} {\bibfnamefont {S.}~\bibnamefont
  {Reyntjens}}\ and\ \bibinfo {author} {\bibfnamefont {R.}~\bibnamefont
  {Puers}},\ }\href@noop {} {\bibfield  {journal} {\bibinfo  {journal} {Journal
  of Micromechanics and Microengineering}\ }\textbf {\bibinfo {volume} {11}},\
  \bibinfo {pages} {287} (\bibinfo {year} {2001})}\BibitemShut {NoStop}%
\bibitem [{\citenamefont {Tseng}\ \emph {et~al.}(2003)\citenamefont {Tseng},
  \citenamefont {Chen}, \citenamefont {Chen},\ and\ \citenamefont
  {Ma}}]{E_beam_lit}%
  \BibitemOpen
  \bibfield  {author} {\bibinfo {author} {\bibfnamefont {A.}~\bibnamefont
  {Tseng}}, \bibinfo {author} {\bibfnamefont {K.}~\bibnamefont {Chen}},
  \bibinfo {author} {\bibfnamefont {C.}~\bibnamefont {Chen}}, \ and\ \bibinfo
  {author} {\bibfnamefont {K.}~\bibnamefont {Ma}},\ }\href {\doibase
  10.1109/TEPM.2003.817714} {\bibfield  {journal} {\bibinfo  {journal}
  {Electronics Packaging Manufacturing, IEEE Transactions on}\ }\textbf
  {\bibinfo {volume} {26}},\ \bibinfo {pages} {141 } (\bibinfo {year}
  {2003})}\BibitemShut {NoStop}%
\bibitem [{\citenamefont {Galisteo-López}\ \emph {et~al.}(2011)\citenamefont
  {Galisteo-López}, \citenamefont {Ibisate}, \citenamefont {Sapienza},
  \citenamefont {Froufe-Pérez}, \citenamefont {Blanco},\ and\ \citenamefont
  {López}}]{ADMA:ADMA201000356}%
  \BibitemOpen
  \bibfield  {author} {\bibinfo {author} {\bibfnamefont {J.~F.}\ \bibnamefont
  {Galisteo-López}}, \bibinfo {author} {\bibfnamefont {M.}~\bibnamefont
  {Ibisate}}, \bibinfo {author} {\bibfnamefont {R.}~\bibnamefont {Sapienza}},
  \bibinfo {author} {\bibfnamefont {L.~S.}\ \bibnamefont {Froufe-Pérez}},
  \bibinfo {author} {\bibfnamefont {l.}~\bibnamefont {Blanco}}, \ and\ \bibinfo
  {author} {\bibfnamefont {C.}~\bibnamefont {López}},\ }\href@noop {}
  {\bibfield  {journal} {\bibinfo  {journal} {Advanced Materials}\ }\textbf
  {\bibinfo {volume} {23}},\ \bibinfo {pages} {30} (\bibinfo {year}
  {2011})}\BibitemShut {NoStop}%
\bibitem [{\citenamefont {López}(2006)}]{1464-4258-8-5-R01}%
  \BibitemOpen
  \bibfield  {author} {\bibinfo {author} {\bibfnamefont {C.}~\bibnamefont
  {López}},\ }\href@noop {} {\bibfield  {journal} {\bibinfo  {journal} {Journal
  of Optics A: Pure and Applied Optics}\ }\textbf {\bibinfo {volume} {8}},\
  \bibinfo {pages} {R1} (\bibinfo {year} {2006})}\BibitemShut {NoStop}%
\bibitem [{\citenamefont {Sapienza}\ \emph {et~al.}(2011)\citenamefont
  {Sapienza}, \citenamefont {Leonetti}, \citenamefont {Froufe-P\'erez},
  \citenamefont {Galisteo-L\'opez}, \citenamefont {Conti},\ and\ \citenamefont
  {L\'opez}}]{PhysRevA.83.023801}%
  \BibitemOpen
  \bibfield  {author} {\bibinfo {author} {\bibfnamefont {R.}~\bibnamefont
  {Sapienza}}, \bibinfo {author} {\bibfnamefont {M.}~\bibnamefont {Leonetti}},
  \bibinfo {author} {\bibfnamefont {L.~S.}\ \bibnamefont {Froufe-P\'erez}},
  \bibinfo {author} {\bibfnamefont {J.~F.}\ \bibnamefont {Galisteo-L\'opez}},
  \bibinfo {author} {\bibfnamefont {C.}~\bibnamefont {Conti}}, \ and\ \bibinfo
  {author} {\bibfnamefont {C.}~\bibnamefont {L\'opez}},\ }\href@noop {}
  {\bibfield  {journal} {\bibinfo  {journal} {Phys. Rev. A}\ }\textbf {\bibinfo
  {volume} {83}},\ \bibinfo {pages} {023801} (\bibinfo {year}
  {2011})}\BibitemShut {NoStop}%
\bibitem [{\citenamefont {Garcia}\ \emph {et~al.}(2010)\citenamefont {Garcia},
  \citenamefont {Sapienza},\ and\ \citenamefont {López}}]{ADMA:ADMA200900827}%
  \BibitemOpen
  \bibfield  {author} {\bibinfo {author} {\bibfnamefont {P.~D.}\ \bibnamefont
  {Garcia}}, \bibinfo {author} {\bibfnamefont {R.}~\bibnamefont {Sapienza}}, \
  and\ \bibinfo {author} {\bibfnamefont {C.}~\bibnamefont {López}},\
  }\href@noop {} {\bibfield  {journal} {\bibinfo  {journal} {Advanced
  Materials}\ }\textbf {\bibinfo {volume} {22}},\ \bibinfo {pages} {12}
  (\bibinfo {year} {2010})}\BibitemShut {NoStop}%
\bibitem [{\citenamefont {Bertolotti}\ \emph {et~al.}(2010)\citenamefont
  {Bertolotti}, \citenamefont {Vynck}, \citenamefont {Pattelli}, \citenamefont
  {Barthelemy}, \citenamefont {Lepri},\ and\ \citenamefont
  {Wiersma}}]{ADFM:ADFM200902008}%
  \BibitemOpen
  \bibfield  {author} {\bibinfo {author} {\bibfnamefont {J.}~\bibnamefont
  {Bertolotti}}, \bibinfo {author} {\bibfnamefont {K.}~\bibnamefont {Vynck}},
  \bibinfo {author} {\bibfnamefont {L.}~\bibnamefont {Pattelli}}, \bibinfo
  {author} {\bibfnamefont {P.}~\bibnamefont {Barthelemy}}, \bibinfo {author}
  {\bibfnamefont {S.}~\bibnamefont {Lepri}}, \ and\ \bibinfo {author}
  {\bibfnamefont {D.~S.}\ \bibnamefont {Wiersma}},\ }\href@noop {} {\bibfield
  {journal} {\bibinfo  {journal} {Advanced Functional Materials}\ }\textbf
  {\bibinfo {volume} {20}},\ \bibinfo {pages} {965} (\bibinfo {year}
  {2010})}\BibitemShut {NoStop}%
\bibitem [{\citenamefont {Barthelemy}\ \emph {et~al.}(2008)\citenamefont
  {Barthelemy}, \citenamefont {Bertolotti},\ and\ \citenamefont
  {Wiersma}}]{Barthelemy2008}%
  \BibitemOpen
  \bibfield  {author} {\bibinfo {author} {\bibfnamefont {P.}~\bibnamefont
  {Barthelemy}}, \bibinfo {author} {\bibfnamefont {J.}~\bibnamefont
  {Bertolotti}}, \ and\ \bibinfo {author} {\bibfnamefont {D.~S.}\ \bibnamefont
  {Wiersma}},\ }\href@noop {} {\bibfield  {journal} {\bibinfo  {journal}
  {Nature}\ }\textbf {\bibinfo {volume} {453}},\ \bibinfo {pages} {495}
  (\bibinfo {year} {2008})}\BibitemShut {NoStop}%
\bibitem [{\citenamefont {Ambartsumyan}\ \emph {et~al.}(1966)\citenamefont
  {Ambartsumyan}, \citenamefont {Basov}, \citenamefont {Kryukov},\ and\
  \citenamefont {Letokhov}}]{1074123}%
  \BibitemOpen
  \bibfield  {author} {\bibinfo {author} {\bibfnamefont {R.}~\bibnamefont
  {Ambartsumyan}}, \bibinfo {author} {\bibfnamefont {N.}~\bibnamefont {Basov}},
  \bibinfo {author} {\bibfnamefont {P.}~\bibnamefont {Kryukov}}, \ and\
  \bibinfo {author} {\bibfnamefont {V.}~\bibnamefont {Letokhov}},\ }\href@noop
  {} {\bibfield  {journal} {\bibinfo  {journal} {Quantum Electronics, IEEE
  Journal of}\ }\textbf {\bibinfo {volume} {2}},\ \bibinfo {pages} {442 }
  (\bibinfo {year} {1966})}\BibitemShut {NoStop}%
\bibitem [{\citenamefont {Wiersma}(2008)}]{Wiersma_Rew}%
  \BibitemOpen
  \bibfield  {author} {\bibinfo {author} {\bibfnamefont {D.~S.}\ \bibnamefont
  {Wiersma}},\ }\href@noop {} {\bibfield  {journal} {\bibinfo  {journal} {Nat
  Phys}\ }\textbf {\bibinfo {volume} {4}},\ \bibinfo {pages} {359} (\bibinfo
  {year} {2008})}\BibitemShut {NoStop}%
\bibitem [{\citenamefont {Leonetti}\ \emph {et~al.}(2009)\citenamefont
  {Leonetti}, \citenamefont {Sapienza}, \citenamefont {Ibisate}, \citenamefont
  {Conti},\ and\ \citenamefont {L\'{o}pez}}]{Leonetti:09}%
  \BibitemOpen
  \bibfield  {author} {\bibinfo {author} {\bibfnamefont {M.}~\bibnamefont
  {Leonetti}}, \bibinfo {author} {\bibfnamefont {R.}~\bibnamefont {Sapienza}},
  \bibinfo {author} {\bibfnamefont {M.}~\bibnamefont {Ibisate}}, \bibinfo
  {author} {\bibfnamefont {C.}~\bibnamefont {Conti}}, \ and\ \bibinfo {author}
  {\bibfnamefont {C.}~\bibnamefont {L\'{o}pez}},\ }\href@noop {} {\bibfield
  {journal} {\bibinfo  {journal} {Opt. Lett.}\ }\textbf {\bibinfo {volume}
  {34}},\ \bibinfo {pages} {3764} (\bibinfo {year} {2009})}\BibitemShut
  {NoStop}%
\bibitem [{\citenamefont {Wu}\ \emph {et~al.}(2006)\citenamefont {Wu},
  \citenamefont {Fang}, \citenamefont {Yamilov}, \citenamefont {Chabanov},
  \citenamefont {Asatryan}, \citenamefont {Botten},\ and\ \citenamefont
  {Cao}}]{PhysRevA.74.053812}%
  \BibitemOpen
  \bibfield  {author} {\bibinfo {author} {\bibfnamefont {X.}~\bibnamefont
  {Wu}}, \bibinfo {author} {\bibfnamefont {W.}~\bibnamefont {Fang}}, \bibinfo
  {author} {\bibfnamefont {A.}~\bibnamefont {Yamilov}}, \bibinfo {author}
  {\bibfnamefont {A.~A.}\ \bibnamefont {Chabanov}}, \bibinfo {author}
  {\bibfnamefont {A.~A.}\ \bibnamefont {Asatryan}}, \bibinfo {author}
  {\bibfnamefont {L.~C.}\ \bibnamefont {Botten}}, \ and\ \bibinfo {author}
  {\bibfnamefont {H.}~\bibnamefont {Cao}},\ }\href@noop {} {\bibfield
  {journal} {\bibinfo  {journal} {Phys. Rev. A}\ }\textbf {\bibinfo {volume}
  {74}},\ \bibinfo {pages} {053812} (\bibinfo {year} {2006})}\BibitemShut
  {NoStop}%
\bibitem [{\citenamefont {Gottardo}\ \emph {et~al.}(2008)\citenamefont
  {Gottardo}, \citenamefont {Sapienza}, \citenamefont {Garcia}, \citenamefont
  {Blanco}, \citenamefont {Wiersma},\ and\ \citenamefont
  {Lopez}}]{Gottardo2008}%
  \BibitemOpen
  \bibfield  {author} {\bibinfo {author} {\bibfnamefont {S.}~\bibnamefont
  {Gottardo}}, \bibinfo {author} {\bibfnamefont {R.}~\bibnamefont {Sapienza}},
  \bibinfo {author} {\bibfnamefont {P.~D.}\ \bibnamefont {Garcia}}, \bibinfo
  {author} {\bibfnamefont {A.}~\bibnamefont {Blanco}}, \bibinfo {author}
  {\bibfnamefont {D.~S.}\ \bibnamefont {Wiersma}}, \ and\ \bibinfo {author}
  {\bibfnamefont {C.}~\bibnamefont {Lopez}},\ }\href@noop {} {\bibfield
  {journal} {\bibinfo  {journal} {Nat Photon}\ }\textbf {\bibinfo {volume}
  {2}},\ \bibinfo {pages} {429} (\bibinfo {year} {2008})}\BibitemShut {NoStop}%
\bibitem [{\citenamefont {Cao}\ \emph {et~al.}(1999)\citenamefont {Cao},
  \citenamefont {Zhao}, \citenamefont {Ho}, \citenamefont {Seelig},
  \citenamefont {Wang},\ and\ \citenamefont {Chang}}]{PhysRevLett.82.2278}%
  \BibitemOpen
  \bibfield  {author} {\bibinfo {author} {\bibfnamefont {H.}~\bibnamefont
  {Cao}}, \bibinfo {author} {\bibfnamefont {Y.~G.}\ \bibnamefont {Zhao}},
  \bibinfo {author} {\bibfnamefont {S.~T.}\ \bibnamefont {Ho}}, \bibinfo
  {author} {\bibfnamefont {E.~W.}\ \bibnamefont {Seelig}}, \bibinfo {author}
  {\bibfnamefont {Q.~H.}\ \bibnamefont {Wang}}, \ and\ \bibinfo {author}
  {\bibfnamefont {R.~P.~H.}\ \bibnamefont {Chang}},\ }\href@noop {} {\bibfield
  {journal} {\bibinfo  {journal} {Phys. Rev. Lett.}\ }\textbf {\bibinfo
  {volume} {82}},\ \bibinfo {pages} {2278} (\bibinfo {year}
  {1999})}\BibitemShut {NoStop}%
\bibitem [{\citenamefont {Garcia}\ \emph {et~al.}(2009)\citenamefont {Garcia},
  \citenamefont {Ibisate}, \citenamefont {Sapienza}, \citenamefont {Wiersma},\
  and\ \citenamefont {L\'opez}}]{PhysRevA.80.013833}%
  \BibitemOpen
  \bibfield  {author} {\bibinfo {author} {\bibfnamefont {P.~D.}\ \bibnamefont
  {Garcia}}, \bibinfo {author} {\bibfnamefont {M.}~\bibnamefont {Ibisate}},
  \bibinfo {author} {\bibfnamefont {R.}~\bibnamefont {Sapienza}}, \bibinfo
  {author} {\bibfnamefont {D.~S.}\ \bibnamefont {Wiersma}}, \ and\ \bibinfo
  {author} {\bibfnamefont {C.}~\bibnamefont {L\'opez}},\ }\href@noop {}
  {\bibfield  {journal} {\bibinfo  {journal} {Phys. Rev. A}\ }\textbf {\bibinfo
  {volume} {80}},\ \bibinfo {pages} {013833} (\bibinfo {year}
  {2009})}\BibitemShut {NoStop}%
\bibitem [{\citenamefont {Fallert}\ \emph {et~al.}(2009)\citenamefont
  {Fallert}, \citenamefont {Dietz}, \citenamefont {Sartor}, \citenamefont
  {Schneider}, \citenamefont {Klingshirn},\ and\ \citenamefont
  {Kalt}}]{Fallert_coexistence_nature}%
  \BibitemOpen
  \bibfield  {author} {\bibinfo {author} {\bibfnamefont {J.}~\bibnamefont
  {Fallert}}, \bibinfo {author} {\bibfnamefont {R.~J.~B.}\ \bibnamefont
  {Dietz}}, \bibinfo {author} {\bibfnamefont {J.}~\bibnamefont {Sartor}},
  \bibinfo {author} {\bibfnamefont {D.}~\bibnamefont {Schneider}}, \bibinfo
  {author} {\bibfnamefont {C.}~\bibnamefont {Klingshirn}}, \ and\ \bibinfo
  {author} {\bibfnamefont {H.}~\bibnamefont {Kalt}},\ }\href@noop {} {\bibfield
   {journal} {\bibinfo  {journal} {Nat Photon}\ }\textbf {\bibinfo {volume}
  {3}},\ \bibinfo {pages} {279} (\bibinfo {year} {2009})}\BibitemShut {NoStop}%
\bibitem [{\citenamefont {van~der Molen}\ \emph {et~al.}(2007)\citenamefont
  {van~der Molen}, \citenamefont {Tjerkstra}, \citenamefont {Mosk},\ and\
  \citenamefont {Lagendijk}}]{PhysRevLett.98.143901}%
  \BibitemOpen
  \bibfield  {author} {\bibinfo {author} {\bibfnamefont {K.~L.}\ \bibnamefont
  {van~der Molen}}, \bibinfo {author} {\bibfnamefont {R.~W.}\ \bibnamefont
  {Tjerkstra}}, \bibinfo {author} {\bibfnamefont {A.~P.}\ \bibnamefont {Mosk}},
  \ and\ \bibinfo {author} {\bibfnamefont {A.}~\bibnamefont {Lagendijk}},\
  }\href@noop {} {\bibfield  {journal} {\bibinfo  {journal} {Phys. Rev. Lett.}\
  }\textbf {\bibinfo {volume} {98}},\ \bibinfo {pages} {143901} (\bibinfo
  {year} {2007})}\BibitemShut {NoStop}%
\bibitem [{\citenamefont {Leonetti}\ \emph {et~al.}(2011)\citenamefont
  {Leonetti}, \citenamefont {Conti},\ and\ \citenamefont
  {Lopez}}]{Leonetti2011}%
  \BibitemOpen
  \bibfield  {author} {\bibinfo {author} {\bibfnamefont {M.}~\bibnamefont
  {Leonetti}}, \bibinfo {author} {\bibfnamefont {C.}~\bibnamefont {Conti}}, \
  and\ \bibinfo {author} {\bibfnamefont {C.}~\bibnamefont {Lopez}},\
  }\href@noop {} {\bibfield  {journal} {\bibinfo  {journal} {Nat Photon}\
  }\textbf {\bibinfo {volume} {5}},\ \bibinfo {pages} {615} (\bibinfo {year}
  {2011})}\BibitemShut {NoStop}%
\bibitem [{\citenamefont {C.J.~Brinker}(1990)}]{Book_1}%
  \BibitemOpen
  \bibfield  {author} {\bibinfo {author} {\bibfnamefont {G.~S.}\ \bibnamefont
  {C.J.~Brinker}},\ }\href@noop {} {\emph {\bibinfo {title} {Sol Gel
  Science}}}\ (\bibinfo  {publisher} {Accademic Press},\ \bibinfo {address}
  {London},\ \bibinfo {year} {1990})\BibitemShut {NoStop}%
\bibitem [{\citenamefont {Chadwick}\ \emph {et~al.}(2002)\citenamefont
  {Chadwick}, \citenamefont {Goodwin}, \citenamefont {Lawson}, \citenamefont
  {Mills},\ and\ \citenamefont {Vincent}}]{Chadwick2002229}%
  \BibitemOpen
  \bibfield  {author} {\bibinfo {author} {\bibfnamefont {M.}~\bibnamefont
  {Chadwick}}, \bibinfo {author} {\bibfnamefont {J.}~\bibnamefont {Goodwin}},
  \bibinfo {author} {\bibfnamefont {E.}~\bibnamefont {Lawson}}, \bibinfo
  {author} {\bibfnamefont {P.}~\bibnamefont {Mills}}, \ and\ \bibinfo {author}
  {\bibfnamefont {B.}~\bibnamefont {Vincent}},\ }\href@noop {} {\bibfield
  {journal} {\bibinfo  {journal} {Colloids and Surfaces A: Physicochemical and
  Engineering Aspects}\ }\textbf {\bibinfo {volume} {203}},\ \bibinfo {pages}
  {229 } (\bibinfo {year} {2002})}\BibitemShut {NoStop}%
\bibitem [{\citenamefont {Leonetti}\ and\ \citenamefont
  {L\'{o}pez}(2011)}]{Leonetti:11}%
  \BibitemOpen
  \bibfield  {author} {\bibinfo {author} {\bibfnamefont {M.}~\bibnamefont
  {Leonetti}}\ and\ \bibinfo {author} {\bibfnamefont {C.}~\bibnamefont
  {L\'{o}pez}},\ }\href@noop {} {\bibfield  {journal} {\bibinfo  {journal}
  {Opt. Lett.}\ }\textbf {\bibinfo {volume} {36}},\ \bibinfo {pages} {2824}
  (\bibinfo {year} {2011})}\BibitemShut {NoStop}%
\bibitem [{\citenamefont {Leonetti}\ \emph {et~al.}(2012)\citenamefont
  {Leonetti}, \citenamefont {Conti},\ and\ \citenamefont
  {L\'opez}}]{Leonetti_pra2011}%
  \BibitemOpen
  \bibfield  {author} {\bibinfo {author} {\bibfnamefont {M.}~\bibnamefont
  {Leonetti}}, \bibinfo {author} {\bibfnamefont {C.}~\bibnamefont {Conti}}, \
  and\ \bibinfo {author} {\bibfnamefont {C.}~\bibnamefont {L\'opez}},\
  }\href@noop {} {\bibfield  {journal} {\bibinfo  {journal} {Phys. Rev. A}\
  }\textbf {\bibinfo {volume} {85}},\ \bibinfo {pages} {043841} (\bibinfo
  {year} {2012})}\BibitemShut {NoStop}%
\end{thebibliography}%

\end{document}